\documentclass[reprint,amsmath,amssymb]{elsarticle}

\usepackage{graphicx}
\usepackage{subfig}
\graphicspath{{./Immagini/}}
\usepackage{dcolumn}
\usepackage{bm}
\usepackage{hyperref}
\usepackage[mathlines]{lineno}
\usepackage[export]{adjustbox}
\usepackage{color}
\usepackage{tabularx,array,booktabs}
\usepackage{caption}
\usepackage{mathtools}
\usepackage{amssymb}

\newcommand{\Om}{\Omega}
\newcommand{\om}{\omega}
\newcommand{\X}{\boldsymbol{X}}
\newcommand{\x}{\boldsymbol{x}}

\newcommand{\F}{\boldsymbol{F}}
\newcommand{\V}{\boldsymbol{V}}
\newcommand{\R}{\boldsymbol{R}}
\newcommand{\g}{\boldsymbol{g}}
\newcommand{\G}{\boldsymbol{G}}
\newcommand{\e}{\boldsymbol{e}}
\newcommand{\E}{\boldsymbol{E}}

\begin{document}

\begin{frontmatter}

\title{Acoustic scattering reduction of elliptical targets via pentamode near-cloaking based on transformation acoustics in elliptic coordinates}

\author[polimi]{Davide Enrico Quadrelli\fnref{myfootnote}}
\author[polimi]{Gabriele Cazzulani}
\author[marina]{Simone La Riviera}
\author[polimi]{Francesco Braghin}
\fntext[myfootnote]{davideenrico.quadrelli@polimi.it}

\address[polimi]{Department of Mechanical Engineering, Politecnico di Milano, via La Masa 1, 20156 Milano, Italy}
\address[marina]{Stato Maggiore della Marina, 5$^o$ Reparto Sommergibili, Piazza della Marina 4, 00196 Roma}
\date{\today}

\begin{abstract}
Cloaks for underwater applications designed for actual submarine acoustic stealth are still far from the technological advancement needed for being put in practice. Several challenges are to be overcame such as dealing with weight or non-axisymmetric shapes. In this paper, we introduce the use of elliptical coordinates to define quasi-symmetric transformations to retrieve the material properties of pentamode cloaks for elliptical shaped targets, along with a quantifiable approximation introduced by the rotation tensor being different from the identity. This is done analitically adopting transformation theory, in an attempt to generalize the usual approach for axisymmetric cloaks, with the aim of dealing with shapes closer to those of the actual cross section of a submarine. With respect to existing techniques for dealing with arbitrarily shaped pentamode cloaks, the introduced technique allows for a priori control on the principal directions of anisotropy and for enlarged design space in terms of possible combinations of material property distributions for the same geometry of the problem.

\end{abstract}


\end{frontmatter}
\section{Introduction}
As a powerful tool for acoustic wave control, Transformation Acoustics has attracted considerable interest over the past decade, primarily because of its association with acoustic cloaking \cite{chen2010acoustic, norris2015acoustic}. The development of perfect acoustic cloaking devices could in principle have a major impact on underwater acoustics, where it could be used to avoid detection of submarines by active sonars \cite{audoly2016perspectives}. However, whether such technology can actually be implemented in real-world applications remains a challenge with several open questions.

Soon after the appearance of the first papers dealing with cloaking in acoustics \cite{cummer2007one, chen2007acoustic}, Norris \cite{Norris2008, Norris2009} pointed to the non-uniqueness of the material properties which the cloak, as a fundamental difference between the acoustic case and the optical case from which the theory was originally initiated \cite{Pendry2006}. In this context, he also pointed out that the cloaks derived directly by analogy with the original theory, which exploit an anisotropic distribution of density, are characterized by a total mass tending towards infinite. This phenomenon, referred to as \textit{mass catastrophe}, makes such \textit{inertial cloaks} (IC) impractical for actual implementations, since the ideal underwater cloak would obviously not affect the buoyancy of the submarine being concealed. 
At the other end of the spectrum of practicable material distributions, cloaks comprising pure \textit{pentamode materials} (PM) \cite{milton1995elasticity} do not suffer from such problem; more than that, they can be implemented adopting solid metafluids. This makes them more suitable for actual implementation with respect to the anisotropic density metafluids, that are in principle obtainable with a layered distribution of conventional fluids \cite{torrent2008acoustic, cheng2008multilayer}. In contrast, the peculiar degenerate stiffness tensor characteristic of PM materials can be achieved by structural optimization of latticed microstructures driven by long wavelength homogenization \cite{layman2013highly, chen2015latticed, Kadic2012, Kadic2013, kadic2014pentamode}. Based on this idea, Yi Chen et al. \cite{chen2017broadband} have shown promising experimental results for a broadband underwater cloak, obtained by optimization of a layered pentamode-inspired microstructure. Recently \cite{chen2019influences}, they also analyzed the impact of the non-zero shear modulus inherent in actual cloaks and showed how sensitive the performance is to different boundary conditions imposed at the inner side of the cloak.

A major challenge in designing practical cloaking devices is related to the shape of the target: indeed, the vast majority of the literature deals only with cases such as the spherical or axisymmetric cloaks, which can be easily treated in an analytical way. Moreover, most of the works dealing with arbitrarily shaped acoustic cloaks refer to the case of IC \cite{li2012homogeneous, li2014two, li2018non, li2019two} and end up in designs involving complicated distributions of layered fluids that would hardly be used in practical applications. \\
The only paper dealing with arbitrarily shaped PM cloaks \cite{chen2016design} proposes a numerical method for obtaining quasi-symmetric transformation gradients to design approximate cloaks. 
This being a very powerful and general tool that can be applied to
any shape of target without restriction, there appears nonetheless to be no a priori control over the distribution of the principal directions of anisotropy within the cloak. This can potentially lead to material distributions that would be not trivial to implement with a microstructure. Moreover, the numerical problem has a unique solution once the geometry and the degree of non-ideality of the cloak are specified, in contrast to the usual analytical solutions which in principle admit an infinite number of solutions, since there is an infinite set of transformations by which the undeformed region is mapped onto the deformed one. Thus, the greater freedom offered by the numerical method in terms the ability to deal with arbitrary shapes, is paid for with a restricted design space in terms of material distributions.

In this paper, we instead focus the attention on the "ellipse in ellipse" cloaking problem and introduce an analytical approach based on the use of the orthogonal elliptic coordinate system, leading to quasi-symmetric transformations that can be used to systematically tackle the design pure pentamode cloaks. Along with this, the maximum and mean local rotation inside the cloak are used as metrics of the non-ideality of the considered design. The idea is to introduce a technique that stands between the exact analytical solution, which is easy to handle for a limited set of shapes, and the numerical approach, with the aim of providing a tool that could facilitate the design of cloaking devices for elliptically shaped targets, which are of interest for underwater applications. Indeed, the introduced set of transformations features a left stretch tensor showing everywhere within the cloak principal directions aligned with local coordinate lines, i.e. normal and tangential to confocal ellipses. This facilitates the construction of the microstructure of the cloak: once the lattice is designed and optimized in rectangular Cartesian coordinates, it can then be "accomodated" around the target by adopting an analytic deformation that automatically aligns the principal directions of anisotropy along the coordinate lines. Moreover, as mentioned, the analytical approach allows infinitely many designs for a given assigned geometry, resulting in an increase of the design space for a given configuration. 

The work is organised as follows: in the next section, the transformation in elliptic coordinates is presented and discussed. A measure of the local rotation introduced by the quasi-symmetric transformation is defined, to quantify the approximation introduced by considering it as a pure stretching. A set of special maps are given as examples such as to obtain constant PM stiffness properties within the cloak, or bulk moduli varying according to power laws. Numerical examples of the performance of the cloak are given to verify the reduction of scattered power achievable even with high approximation introduced. Before conclusions are drawn, Section 3 also discusses the problem of accommodating a microstructure designed in rectangular coordinates around the target adopting conformal maps, such that the principal directions of anisotropy are aligned as required.

\section{Transformation in elliptic coordinates}
\begin{figure*}[h]
 \centering
 \includegraphics[width=0.8\textwidth,angle=0] {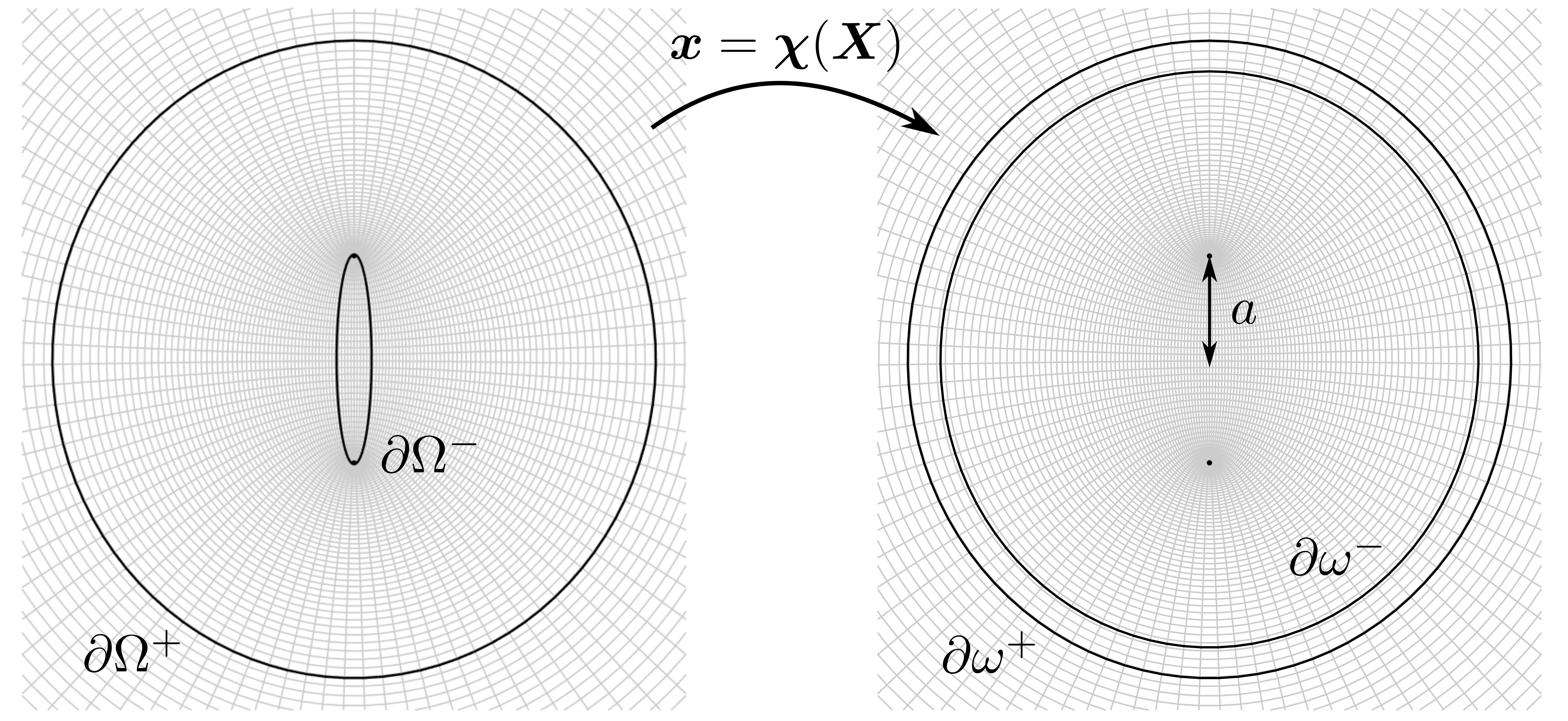}
 \caption{Graphical representation of the transformation $\x=\boldsymbol{\chi}(\X)$ from the undeformed domain $\Omega$ (\textit{left}), to the virtual deformed one $\omega$ (\textit{right}). Coordinate lines for the elliptic coordinates system are superposed in grey color. Constant $\mu$ coordinate lines correspond to confocal ellipses, allowing easy handling of elliptic boundaries. The location of the two foci is $\pm a$ and is indicated by $a$.}
\label{Figure1}
\end{figure*}
The key idea behind Transformation Acoustics is the definition of  a point-wise transformation $\boldsymbol{\chi}$ from  $\X \in \Omega$ to $\x=\boldsymbol{\chi}(\X)\in \omega$ between points in the so called virtual undeformed domain $\Omega$ to points in the deformed virtual one $\omega$ (Figure \ref{Figure1}).
Cloaking is achieved when the map is the identity on the outer boundaries of the domains $\partial \Omega^+=\partial \omega^+$, and the inner boundary of the cloak $\partial \omega^-$ is mapped through the inverse transformation to a small surface with vanishing scattering cross section. Adopting the language of finite deformations, the deformation gradient $\F$ has determinant $J=\operatorname{det}(\F)$ equal to the ratio of volume elements in the two configurations and can be decomposed according to $\F=\V \R$ where $\R$ is an orthogonal tensor ($\operatorname{det}(\R)=1$, $\R\R^T=\R^T\R=\boldsymbol{I}$) representing local rotations while the symmetric and positive definite $\V\in\operatorname{Sym}^+$ represents local stretches. It has been previously shown \cite{Norris2008} \cite{chen2016design} that the fundamental requirement for obtaining a pure pentamode cloak is that the transformation is a pure stretch, i.e. $\R=\boldsymbol{I}$ and $\F=\V$ is symmetric.

\begin{figure*}[t]
 \centering
 \includegraphics[width=\textwidth,angle=0] {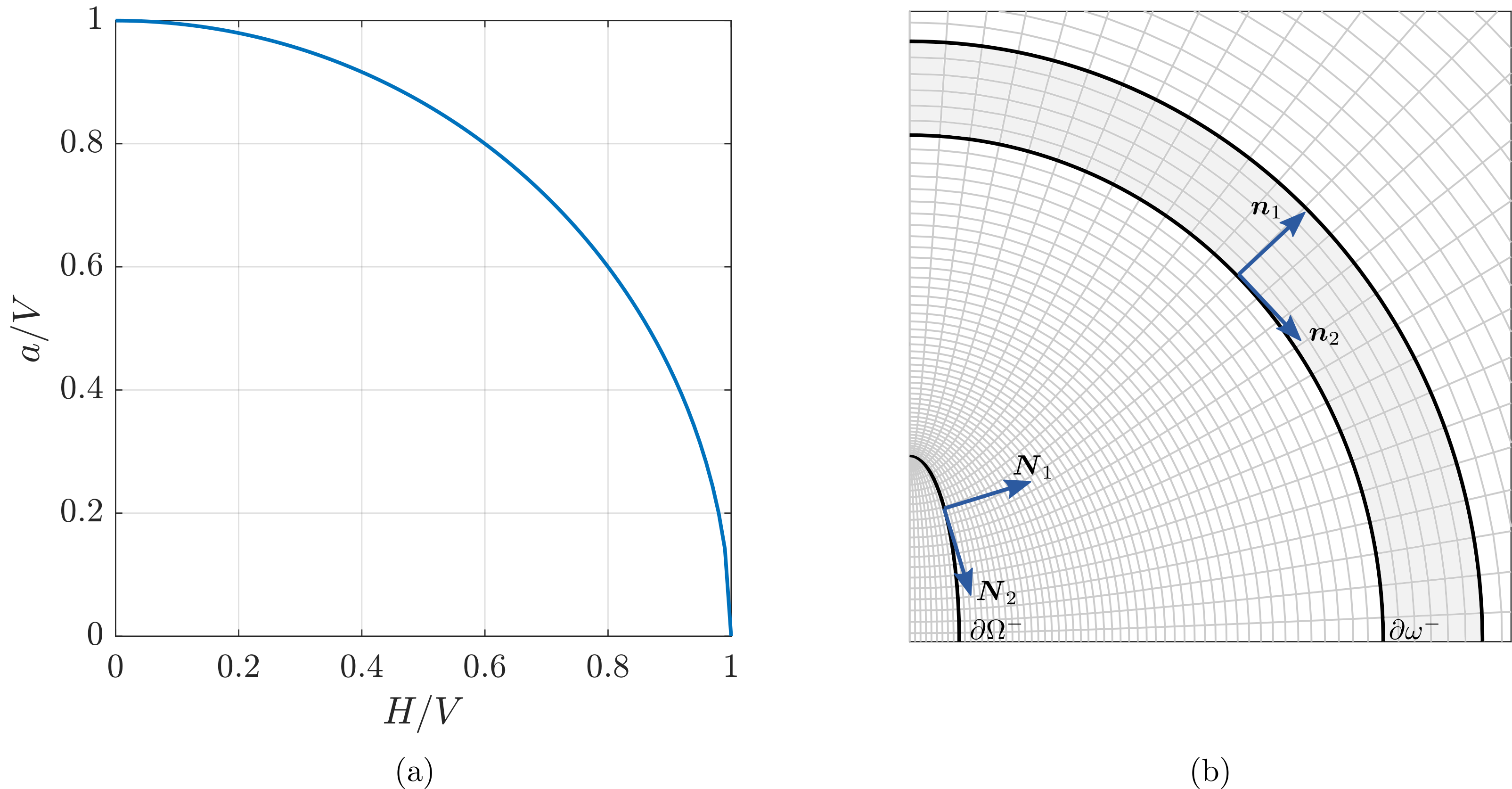}
 \caption{Figure \ref{Figure2}(a): Ratio of the half focal distance to the vertical semi-axis as a function of the ellipticity of the target to be cloaked, expressed as the ratio of the horizontal to the vertical semi-axis. Figure \ref{Figure2}(b) Pictorial representation of the rotation of coordinate basis vectors after transformation for a point located on the inner boundary $\partial \Omega^-$}
\label{Figure2}
\end{figure*}

Let us briefly recall the relationship between cartesian coordinates and elliptic coordinates:
\begin{equation}
\begin{cases}
x=a \sinh \mu \sin \nu\\
y=a \cosh \mu \cos \nu
\end{cases}
\end{equation}
For constant $\mu$, coordinate lines are ellipses with vertical semi-axis equal to $a \cosh \mu$ and horizontal semi-axis equal to $a \sinh \mu$, as shown in Figure \ref{Figure1}. The location of the two foci is $y=\pm a$ for any choice of $\mu$. For constant values of $\nu$, hyperbole are obtained with focal distance coincident with that of the ellipses.
As a consequence, the use of elliptic coordinates allows to handle easily transformations between domains whose boundaries are ellipses.
Indeed, referring to Figure \ref{Figure1}, undeformed and deformed domains can be described respectively as:
\begin{equation}
\Om = \left\{ \X : (X, Y)=(a \sinh(R) \sin \Theta, a \cosh(R) \cos\Theta),  R \in [R_1,R_3], \Theta \in [0,2\pi] \right\}
\end{equation}
and
\begin{equation}
\om = \left\{ \x : (x,y) =(a \sinh(r) \sin \theta, a \cosh(r)\cos \theta), r \in [R_2,R_3], \theta \in [0,2\pi] \right\}
\end{equation}
having adopted capital letters for coordinates in the undeformed configuration and plain letters for the deformed one. The use of $R,r$ and $\Theta, \theta$ is introduced to enforce intuitive similarity between elliptic and polar coordinates. Indeed $\Theta, \theta$ span between $0$ and $2\pi$ and behave like the polar angle, while $R,r$ are non negative variables that play a role similar to a radial coordinate. Polar coordinates can indeed be seen as a limit where the focal distance $2a$ tends to zero. The cloak is here obtained in analogy with the usual approach for axisymmetric cloaks, considering $0<R_1<R_2<R_3$, where $R_1$ is a small non vanishing value and the ellipse defined by $r=R_2$ corresponds to the outer surface of the object to be concealed. On the other hand, the coordinate line corresponding to $r=R_3$ corresponds to the outer surface of the cloak. It is thus evident at this point that the approach poses a restriction on the shape of the outer boundary $\partial\Omega^+$ and on the shape of the equivalent acoustic target $\partial\Omega^-$: they both have to be confocal to the ellipse describing the shape of the target. Indeed, when the horizontal and vertical semi-axis $H$ and $V$ of the target are specified, then the focal distance is univocally defined by:
\begin{equation}
    \begin{cases}
        a=\displaystyle\frac{H}{\sinh(R_2)}=\displaystyle\frac{V}{\cosh(R_2)}\\
        R_2=\operatorname{atanh}\left(\displaystyle\frac{H}{V}\right)
    \end{cases}
    \label{eqfocal}
\end{equation}
There are thus similarities in this approach to what is done for the usual axisymmetric cloak, where a small circular hole of vanishing scattering cross-section is "enlarged" to occupy the entire area of the object to be cloaked, resulting in radial and tangential stretching of space into a doughnut that occupies the area of the cloak. Nevertheless,there are differences worth pointing out. First, in the undeformed configuration the hole tends to be a linear segment rather than a point when the parameter $R_1$, which controls the apparent acoustic size of the cloaked target, approaches zero. Indeed, $R,r=0$ corresponds to the line connecting the focal points. This in turn leads to the fact that the performance of the cloak is strongly dependent on the direction of incidence. In the limit for $R_1=0$, the line has zero dimension for plane waves propagating in the direction parallel to it, so that in this case perfect cloaking is achieved. In contrast, for plane waves whose wavevector is aligned perpendicularly to it and thus aligned with the minor axis, the cloaked object behaves as if it were a rigid, flat obstacle with dimensions $2a$. This in turn means that the scattering reduction in this latter case has an upper bound set by the shape of the object to be cloaked, that uniquely determines the focal length $2a$, as first indicated by Eq. (\ref{eqfocal}). In particular, as the ratio between minor and major axes of the ellipse becomes smaller, the ratio of the focal length over the vertical semi-axis becomes larger, decreasing the upper limit of scattering reduction. The aforementioned trend is shown in Figure \ref{Figure2}(a), where the location of the focal points $a$, normalized with respect to the major axis, is plotted against the ratio $H/V$, which is used as a measure of the ellipticity of the target.
The direct mapping $\x=\boldsymbol{\chi} (\X)$ at this point can thus be written as:
\begin{equation}
\begin{cases}
r=g(R)\\
\theta=\Theta
\end{cases}
\end{equation}
with the inverse transformation $\X = \boldsymbol{\chi}^{-1}(\x)$ corresponding to:
\begin{equation}
\begin{cases}
R=f(r)\\
\Theta=\theta
\end{cases}
\end{equation}
The deformation gradient reads:
\begin{equation}
\F=\frac{1}{f^\prime(r)} \g_r \otimes \G^R + \g_\theta \otimes \G^{\Theta}
\label{eq_F}
\end{equation}
being $\g_i$ the covariant basis in the deformed configuration, and $\G^i$ the set of contravriant basis vectors in the undeformed one (see \ref{AppeA}). The components of the transpose of the deformation gradient can be found according to \cite{marsden1994mathematical}:
\begin{equation}
    (\F^T)^A_{\;\;\;a}=g_{ab}F^b_{\;\;\;B}G^{AB}
\end{equation}
where $g_{ab}$ and $G^{AB}$ stand for the components of the metric tensors. It follows that:
\begin{equation}
    \F^T =\frac{1}{f^{'}(r)}\frac{\sinh^2(r) + \sin^2 \theta}{\sinh^2(f(r)) + \sin^2 \theta} \G_R \otimes \g^r + \frac{\sinh^2(r) + \sin^2 \theta}{\sinh^2(f(r)) + \sin^2 \theta} \G_\Theta \otimes \g^{\theta}
\end{equation}
 The left stretch tensor can be then found with the definition $\V^2=\F\F^T$:
\begin{equation}
    \V^2 =\frac{1}{f^{'}(r)^2}\frac{\sinh^2(r) + \sin^2 \theta}{\sinh^2(f(r)) + \sin^2 \theta} \g_r \otimes \g^r + \frac{\sinh^2(r) + \sin^2 \theta}{\sinh^2(f(r)) + \sin^2 \theta} \g_\theta \otimes \g^{\theta}
\end{equation}
$\V$ is diagonal and expressed with respect to an orthogonal basis, thus principal directions of stretch can be identified as:
\begin{equation}
    \begin{split}
        \boldsymbol{n}_1=\frac{1}{|\g_r|}\g_r\\
        \boldsymbol{n}_2=\frac{1}{|\g_\theta|}\g_\theta
    \end{split}
\end{equation}
i.e. everywhere tangent to confocal ellipses and hyperbolae. 
Thus:
\begin{equation}
    \V=\lambda_1 \boldsymbol{n}_1 \otimes \boldsymbol{n}_1 + \lambda_2 \boldsymbol{n}_2 \otimes \boldsymbol{n}_2
\end{equation}
Where the principal stretches read:
\begin{equation}
\begin{split}
&\lambda_1=F^r_{\;\;\;R}\frac{|\g_r|}{|\G_R|}=\frac{1}{f^\prime(r)}\frac{\sqrt{\sinh^2(r)+\sin^2 \theta}}{\sqrt{\sinh^2(f(r))+\sin^2 \theta}}\\
&\lambda_2=F^\theta_{\;\;\;\Theta}\frac{|\g_\theta|}{|\G_\Theta|}=\frac{\sqrt{\sinh^2(r)+\sin^2 \theta}}{\sqrt{\sinh^2(f(r))+\sin^2 \theta}}
\end{split}	
\end{equation}
Then:
\begin{equation}
    \F=\V\R=(\lambda_1\boldsymbol{n}_1 \otimes \boldsymbol{n}_1+\lambda_2 \boldsymbol{n} \otimes \boldsymbol{n}_2)(\boldsymbol{n}_1 \otimes \boldsymbol{N}_1+ \boldsymbol{n}_2\otimes \boldsymbol{N}_2)=\lambda_1\boldsymbol{n}_1\otimes\boldsymbol{N}_1+\lambda_2\boldsymbol{n}_2\otimes\boldsymbol{N}_2
    \label{eq_Fgeneral}
\end{equation}
being $\boldsymbol{N}_1$ and $\boldsymbol{N}_2$ the principal directions of the right stretch tensor $\boldsymbol{U}$ according to the alternative polar decomposition $\F=\R\boldsymbol{U}$. Comparing Eq.(\ref{eq_Fgeneral}) with Eq. (\ref{eq_F}) it is possible to find:
\begin{equation}
    \begin{split}
        \boldsymbol{N}_1=\frac{\G^R}{|\G^R|}=\frac{\G_R}{|\G_R|}\\
        \boldsymbol{N}_2=\frac{\G^\Theta}{|\G^\Theta|}=\frac{\G_\Theta}{|\G_\Theta|}
    \end{split}
\end{equation}
from which the rotation tensor can be evaluated as:
\begin{equation}
    \R=\boldsymbol{n}_1\otimes\boldsymbol{N}_1+\boldsymbol{n}_2\otimes\boldsymbol{N}_2=\frac{|\G_R|}{|\g_r|}\g_r\otimes\G^R+\frac{|\G_\Theta|}{|\g_\theta|}\g_\theta\otimes \G^\Theta
    \label{eq_rotation}
\end{equation}
Being, in general, $\boldsymbol{n}_i\neq \boldsymbol{N}_i$ (without summation implied) the transformation is not a pure stretch. Nonetheless, it is possible to show that under certain assumptions the maximum and mean rotation remain bounded to low values and can be neglected. If such condition is met, material properties can be considered to correspond to that of a pure pentamode material:
\begin{equation}
\begin{split}
&\rho=J^{-1}=(\lambda_1\lambda_2)^{-1}=f^\prime(r)\frac{\sinh^2(f(r))+\sin^2 \theta}{\sinh^2(r)+\sin^2 \theta}\\
&K=J=\lambda_1\lambda_2=\frac{1}{f^\prime(r)}\frac{\sinh^2(r)+\sin^2 \theta}{\sinh^2(f(r))+\sin^2 \theta}\\
&\boldsymbol{S}=J^{-1}\V=\frac{\sqrt{\sinh^2(f(r))+\sin^2 \theta}}{\sqrt{\sinh^2(r)+\sin^2 \theta}}\boldsymbol{n}_1 \otimes \boldsymbol{n}_1 + f^\prime(r)\frac{\sqrt{\sinh^2(f(r))+\sin^2 \theta}}{\sqrt{\sinh^2(r)+\sin^2 \theta}} \boldsymbol{n}_2 \otimes \boldsymbol{n}_2\\
&\mathbb{C}=K \boldsymbol{S} \otimes \boldsymbol{S} 
\end{split}
\end{equation}
The angle of rotation $\alpha$ can be evaluated by the scalar product $\boldsymbol{n}_1 \cdot \boldsymbol{N}_1=\cos\alpha$, which gives:
\begin{equation}
    \cos \alpha=\frac{\sin^2(\theta)\cosh(r)\cosh(f(r))+\cos^2(\theta)\sinh(r)\sinh(f(r))}{\sqrt{\sinh^2(r)+\sin^2(\theta)}\sqrt{\sinh^2(f(r))+\sin^2(\theta)}}
\end{equation}
Figure \ref{Figure2}(b) shows the boundaries of the deformed and undeformed domains overlaid with coordinate lines. Unit vectors $\boldsymbol{N}_i$ and $\boldsymbol{n}_i$ are shown for a point along the inner boundaries $\partial \Omega^-$ $\partial \omega^-$, the visualization of which provides a better understanding of the nature of local rotations (the length of the unit vectors is magnified for illustration). Since the first principal direction is everywhere tangent to the hyperbolas, the local rotation follows from the curvature of the hyperbolas near the foci. Since hyperbolas far from the foci tend asymptotically to straight lines, it follows that for transformations where $g(R)>R$ everywhere, the maximum of local rotation is expected for points originally lying on $\partial \Omega^{-}$. Such maximum rotation is thus related both to the parameter $R_1$, which defines the location of $\partial \Omega^{-}$, and to the ratio $H/V$, which gives the shape of the target, on which the ratio $a/V$ depends. The smaller the focal distance with respect to the vertical semi-axis, the smaller the region of space in which hyperbolas are characterized by high curvature, while the higher $R_1$ is, the farther the inner boundary $\partial \Omega^{-}$ is from this region. 

It might appear that satisfying the low rotation condition imposes some significant constraints both on the applicability of the approach, since it is restricted to low ellipticity targets, and on the performance of the solution, since a lower bound on $R_1$ is imposed. In fact, actual submarine shapes are never characterized by extreme differences between major and minor semi-axes, due to symmetry reasons related to the structural resistance of the hull \cite{burcher1995concepts}. On the other hand, very low values of $R_1$ are ruled out in principle for buoyancy reasons: considering the material properties of a pure pentamode cloak, the density is proportional to the reciprocal of the Jacobian, i.e., the remaining Archimedes thrust available to the target after subtracting the weight of the cloak from the total thrust is equal to the weight of the water volume contained in $\partial \Omega^-$. Thus, $R_1 \rightarrow 0$ would ideally correspond to a cloak weight exactly equal to Archimedes total thrust. For this reason, most actual feasible implementations will be characterized by rotations limited to negligible values.
\subsection{Special Transformations for Elliptic Cloak}
\begin{figure*}[t]
 \centering
 \includegraphics[width=0.7\textwidth,angle=0] {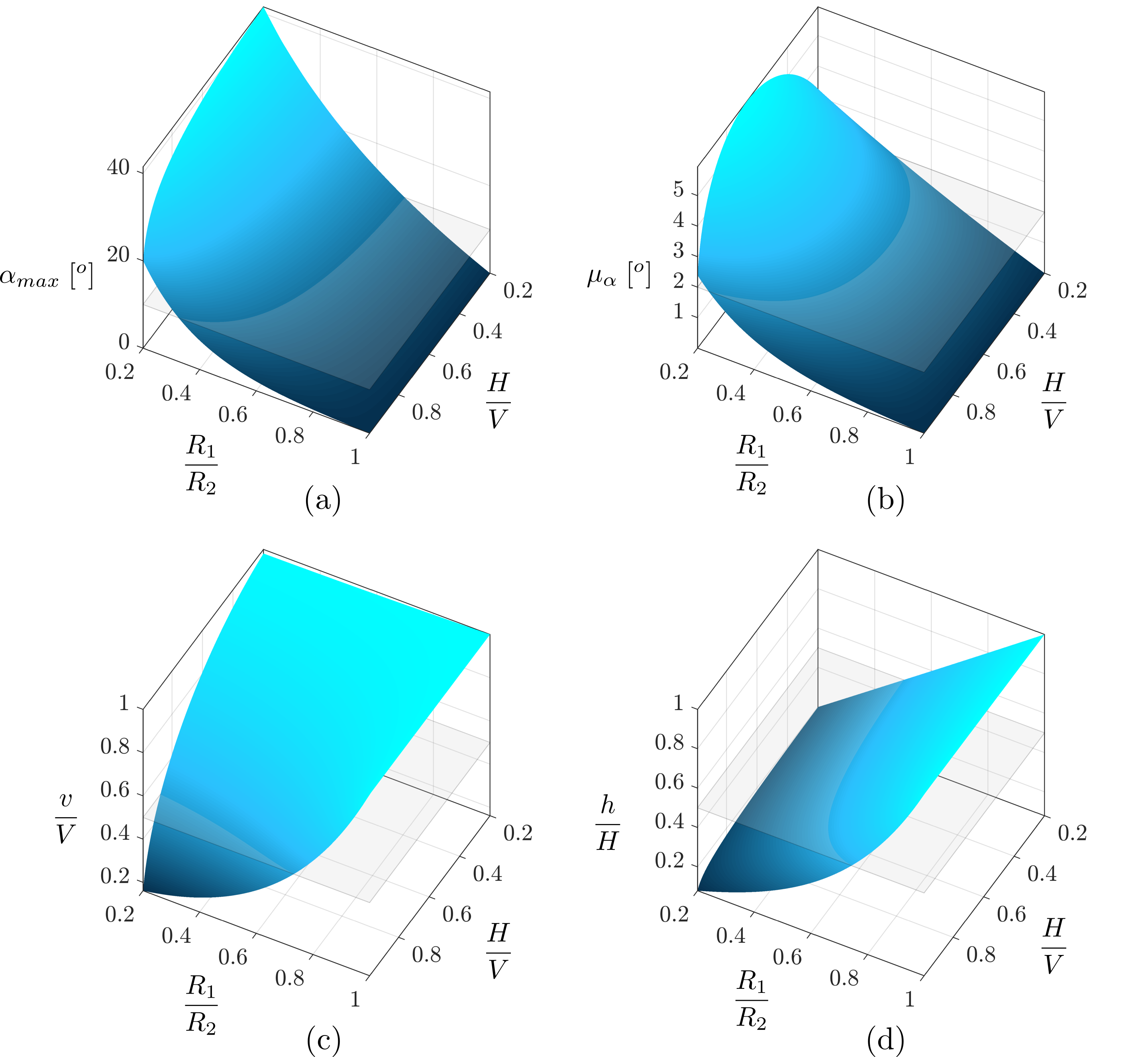}
 \caption{Figure \ref{Figure3}(a): Calculated maximum local rotation for linear transformation when considering combinations of the geometrical parameter $H/V \in [0.2, 1]$ and the design parameter $R_1/R_2 \in [0.2, 1]$. Figure \ref{Figure3}(b): Local rotation averaged over the cloak domain for linear transformation and same range of parameters as in Figure \ref{Figure3}(a). Figure \ref{Figure3}(c): Geometrical reduction of vertical semi-axis. Figure \ref{Figure3}(d) Geometrical reduction of horizontal semi-axis.} 
\label{Figure3}
\end{figure*}
Once the geometry of the problem at hand is specified in terms of $R_1$,$R_2$ and $R_3$, the analytic approach allows infinitely many solutions as the infinitely many possible transformations that can be written between the deformed and undeformed domains. This allows to control the distribution of material properties, giving the possibility to obtain different combinations of density and bulk moduli. Therefore, in the following we give as an example a set of special transformations in elliptic coordinates. It shuould be noted that the scaling factors of the elliptic coordinate system are the same (\ref{AppeA}: $|\G_R|=|\G_\Theta|$, $|\g_r|=|\g_\theta|$), so that the stiffness tensor turns out to depend only on the derivative of the mapping between $R$ and $r$:
\begin{equation}
    \begin{split}
    [C]&=\begin{bmatrix}
	\displaystyle\frac{1}{f^\prime(r)}\frac{|\g_r|}{|\G_R|}\frac{|\G_\Theta|}{|\g_\theta|}	& 1	& 0\\
	1 & f^\prime(r)\displaystyle\frac{|\G_R|}{|\g_r|}\frac{|\g_\theta|}{|\G_\Theta|} & 0\\
	0 & 0 & 0
	\end{bmatrix}= \begin{bmatrix}
	\displaystyle\frac{\lambda_1}{\lambda_2} & 1 & 0 \\
	1 & \displaystyle\frac{\lambda_2}{\lambda_1} & 0\\
	0 & 0 & 0
	\end{bmatrix}=\\
	&=\begin{bmatrix}
    \displaystyle\frac{1}{f^\prime(r)} & 1 & 0 \\
    1 & f^{\prime}(r) & 0\\
    0 & 0 & 0
    \end{bmatrix}
    \end{split}
\end{equation}
This results in ease of tuning of the definition of $f(r)$ upon requirements on the stiffness properties of the cloak.
\paragraph{Transformation for constant bulk moduli}
\begin{figure*}[t]
 \centering
 \includegraphics[width=\textwidth,angle=0] {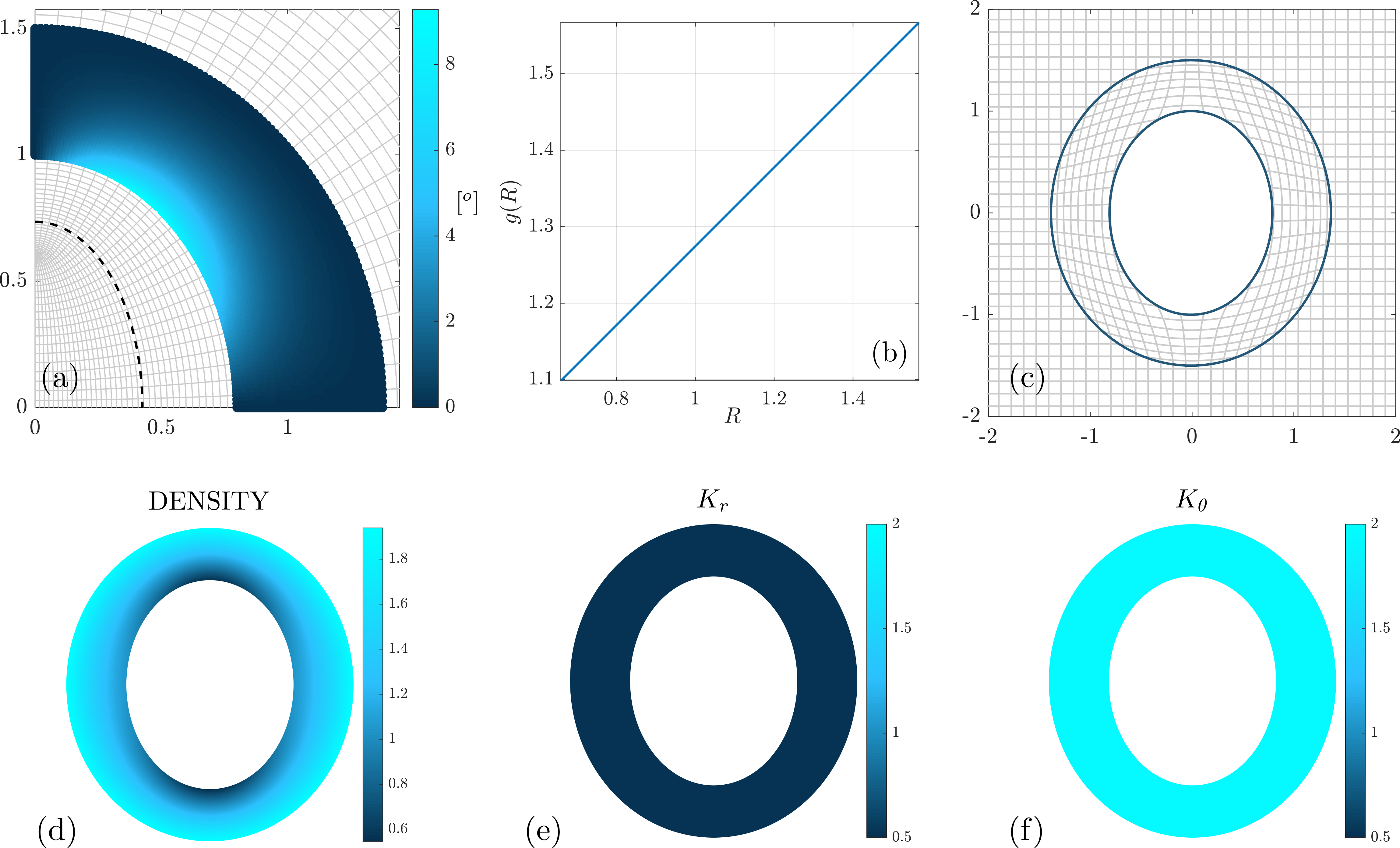}
 \caption{Figure \ref{Figure5}(a): Distribution of local rotation angle $\alpha$ for a linear transformation when $H/V=0.8$, $R_1/R_2=0.6$ and the outer vertical semi-axis of the cloak is selected to be $1.5V$. Figure \ref{Figure5}(b): Direct map $r=g(R)$. Figure \ref{Figure5}(c) Deformation of straight lines inside the cloak when the direct mapping $r=g(R)$ is applied. Figure \ref{Figure5}(d)-(e)-(f) Calculated density and bulk moduli in principal directions. The shown values are normalized on the background fluid's physical properties.}
\label{Figure5}
\end{figure*}
\begin{figure*}[t]
 \centering
 \includegraphics[width=\textwidth,angle=0] {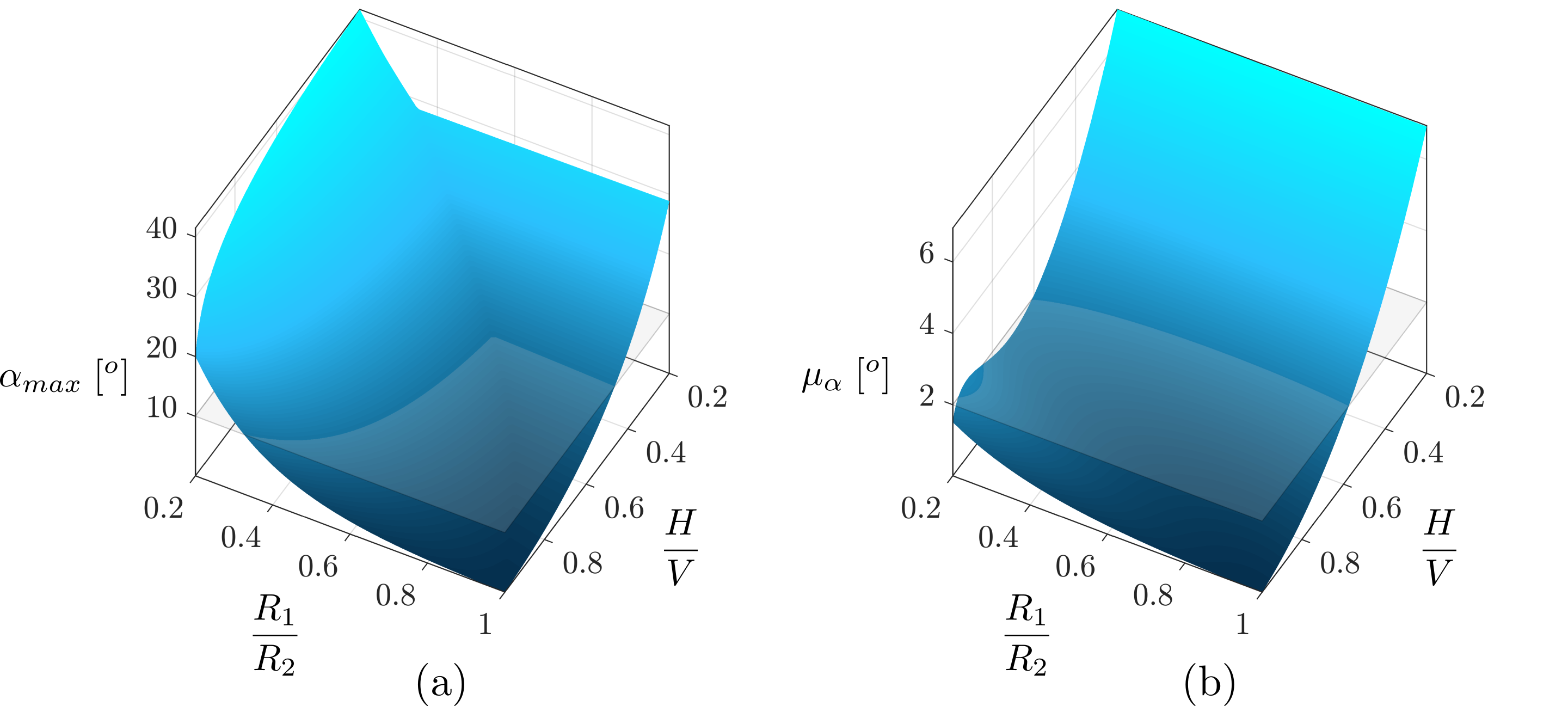}
 \caption{Figure \ref{Figure4}(a): Calculated maximum local rotation for power law transformation when considering combinations of the geometrical parameter $H/V \in [0.2, 1]$ and the design parameter $R_1/R_2 \in [0.2, 1]$. Figure \ref{Figure4}(b): Local rotation averaged over the cloak domain for power law transformation and same range of parameters as in Figure \ref{Figure3}(a).}
\label{Figure4}
\end{figure*}

\begin{figure*}[t]
 \centering
 \includegraphics[width=\textwidth,angle=0] {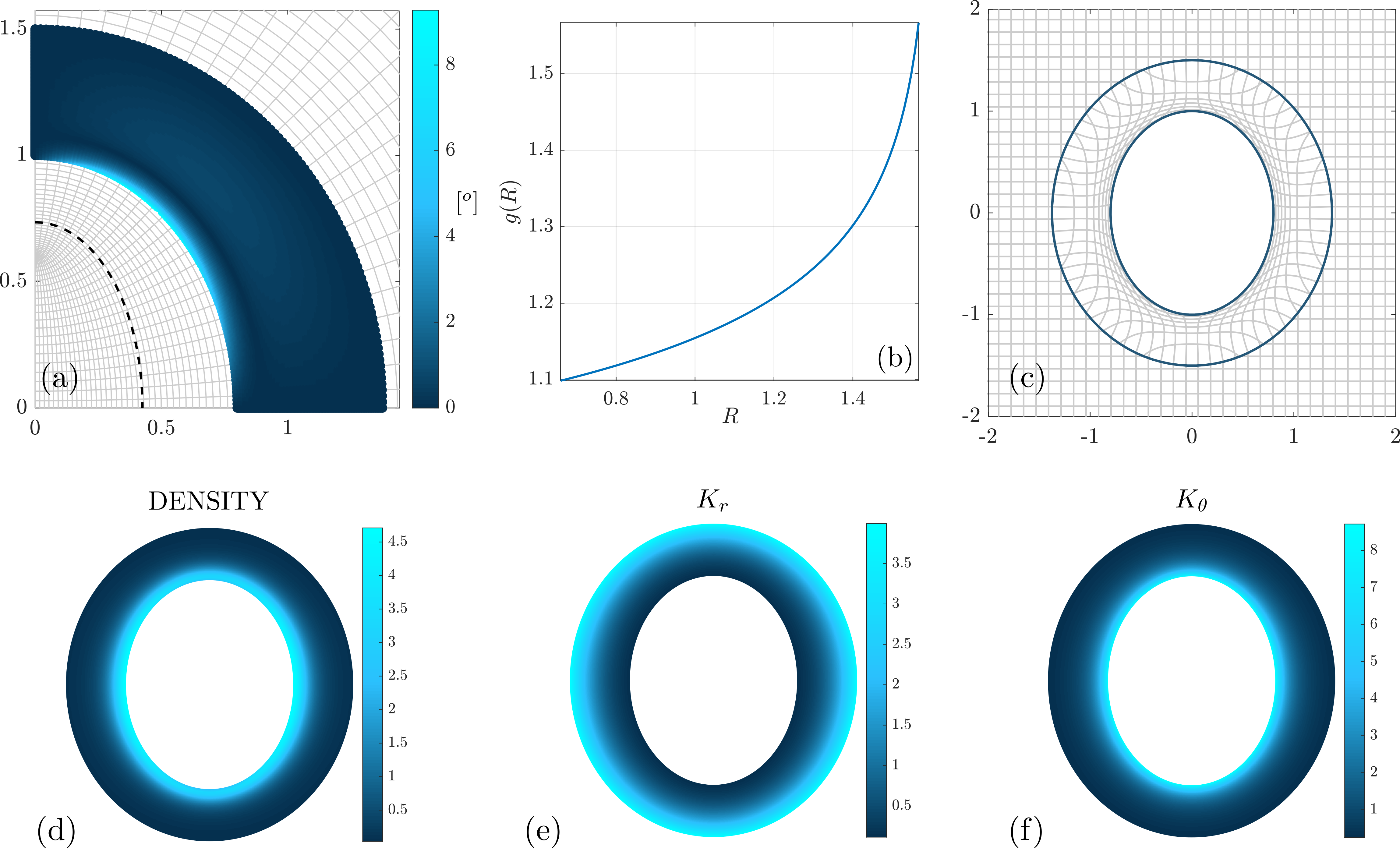}
 \caption{Figure \ref{Figure6}(a): Distribution of local rotation angle $\alpha$ for a power law transformation when $\beta=-10$, $H/V=0.8$, $R_1/R_2=0.6$ and the outer vertical semi-axis of the cloak is selected to be $1.5V$. Figure \ref{Figure6}(b): Direct map $r=g(R)$. Figure \ref{Figure6}(c) Deformation of straight lines inside the cloak when the direct mapping $r=g(R)$ is applied. Figure \ref{Figure6}(d)-(e)-(f) Calculated density and bulk moduli in principal directions. The shown values are normalized on the background fluid's physical properties.}
\label{Figure6}
\end{figure*}
To obtain constant $K_r, K_\theta$ it is sufficient to use a linear transformation subjected to the constrains:
\begin{equation}
    \begin{cases}
    f(R_2)=R_1\\
    f(R_3)=R_3
    \end{cases}
\end{equation}
which results in 
\begin{equation}
    f(r)=\frac{R_3-R_1}{R_3-R_2}(r-R_2)+R_1
\end{equation}
This transformation is interesting for the implementation of microstructures, since the gradient of material properties is related to only one parameter, namely density. The maximum and average rotation within the cloak can be taken as a measure of the introduced non-ideality of the cloak and the approximation introduced in the solution by considering the transformation gradient equal to the left stretch tensor. As mentioned before, such figures of merit depend on both the $H/V$ ratio, which specifies the shape of the target to be cloaked, and the $R1/R2$ ratio. Figure \ref{Figure3}(a) and (b) represent the maximum and average value of $\alpha$, respectively, for combinations of such independent parameters. Each possible combination leads to a different associated geometric reduction $v/V$ and $h/H$, which are shown in Figure \ref{Figure3}(c)-(d), being $v$ and $h$ the vertical and horizontal semi-axes of $\partial \Omega^-$. This provides a better intuition on the equivalent apparent reduction in size of the target: note that while $h$ can take values tending to zero, $v$ must be boundedly larger than $a$. Horizontal planes are introduced for a better visualization of the results.

Figure \ref{Figure5}(a) shows the calculated distribution of rotation $\alpha$ inside the cloak for a target characterized by $V=1$, $H/V=0.8$, and $R_1/R2=0.6$. The outer boundary of the cloak $\partial \omega^+$ is chosen so that its vertical semi-axis is 1.5 times $V$. The dashed line marks the location of $\partial \Omega^-$. Besides, in Figure \ref{Figure5}(b) the direct mapping $g(R)$ is shown, while in Figure \ref{Figure5}(c) straight lines in $\Omega$ are deformed according to the direct transformation for graphical visualization of $\boldsymbol{\chi}$. Finally, Figures \ref{Figure5}(d)-(e)-(f) show the calculated corresponding distributions of density and bulk moduli.
\paragraph{Power law for bulk modulus}
The aim is now to find a constant $K_1$ such that the bulk moduli can be defined as:
\begin{equation}
    \begin{cases}
    K_\theta=K_1\left( \displaystyle\frac{r}{R_2} \right)^\beta\\
    K_r=\displaystyle\frac{1}{K_\theta}
    \end{cases}
\end{equation}
The transformation must satisfy the following differential equation:
\begin{equation}
    f^\prime(r)=K_1\left( \frac{r}{R_2} \right)^\beta 
\end{equation}
subjected to the usual boundary conditions $f(R_2)=R_1$, $f(R_3)=R_3$.
The solution is thus:
\begin{equation}
    f(r)=\frac{K_1}{\beta+1}\left( \frac{r}{R_2} \right)^{\beta+1}R_2+K_2
\end{equation}
with
\begin{equation}
    \begin{split}
        &K_1=\frac{(R_3-R_1)(\beta+1)}{R_2}\left[ \left( \frac{R_3}{R_2} \right)^{\beta+1}-1\right]\\
        &K_2=R_1-\frac{R_2}{\beta+1}K_1
    \end{split}
\end{equation}
For illustration purposes Figure \ref{Figure4} shows the calculated maximum and average values of the local rotation $\alpha$ for combinations of geometric and design parameters $H/v$ and $R_1/R_2$, when the exponent is set to $\beta=-10$. 

The increased freedom in the choice of the transformation given by the analytical approach allows the design to be tailored to the needs of the specific application, e.g., to obtain simpler material distributions to be implemented with a microstructure or to pursue minimization of local rotation. As an example, in Figure \ref{Figure6}, a cloak with the same geometric features assumed for the numerical example shown in Figure \ref{Figure5} is now designed using a power law with $\beta=-10$. As can be seen in Figure \ref{Figure6}(a)-(b)-(c) this particular transformation allows for reduced average rotation within the cloak, due to the smaller tangent of $g(R)$ near $\partial \Omega^-$. On the other hand, one pays for this with a more complicated distribution of density and bulk moduli compared to those obtained with the linear mapping, as shown in Figure \ref{Figure6}(c)-(d)-(e). Note that for this type of transformation, $g(R)$ is not everywhere larger than $R$. Thus, there are points that decrease their distance from the origin. This is the reason why the maximum rotations shown in Figures \ref{Figure3}(a) and \ref{Figure4}(a) are not the same. even though the geometry of the problem is the same.\\

At this point, it is logical to ask whether it is possible to obtain cloaks with constant isotropic density. Considering that the problem is not axisymmetric, it is reasonable to expect that the dependence on $\theta$ in the distribution of anisotropic propagation velocities
\begin{equation}
    v_r=\sqrt{\frac{K_r}{\rho}}; \qquad v_\theta=\sqrt{\frac{K_\theta}{\rho}}
\end{equation}
cannot be eliminated. Since no dependence on $\theta$ is found for the components of the elasticity tensor $\mathbb{C}$, it follows that the density inside the cloak can never be constant, since the dependence of the solution on $\theta$ is required.\\

\begin{figure*}[t]
 \centering
 \includegraphics[width=\textwidth,angle=0] {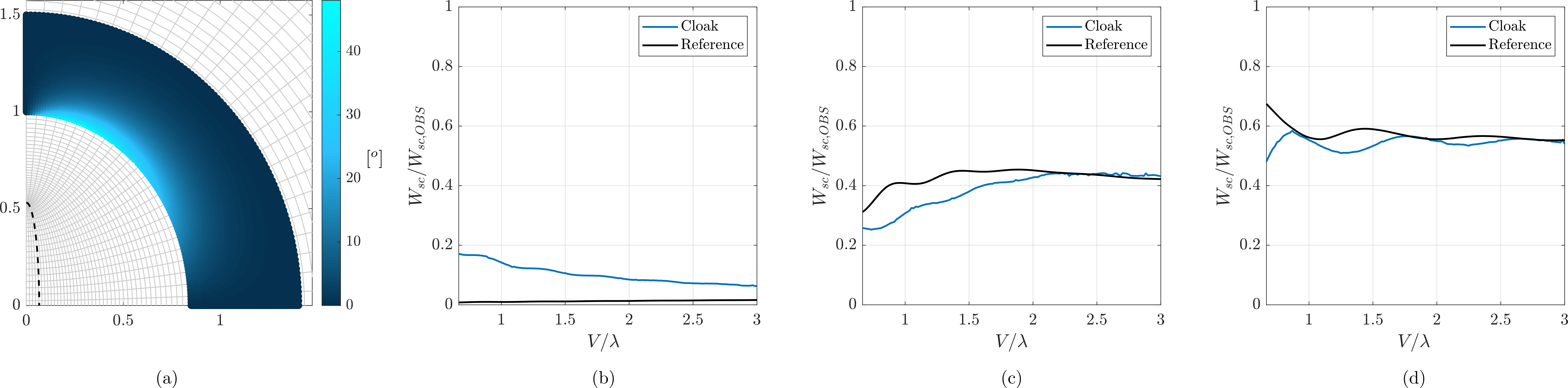}
 \caption{Figure \ref{Figure7}(a) Calculated local rotation angle $\alpha$ for the cloak's performance numerical assessment. The selected parameters are: $H/V=0.85$, $R_1/R_2=0.1$ outer vertical semi-axis is $1.5V$. The transformation is obtained with a linear map. Figure \ref{Figure7}(b) Performance in terms of the scattered power normalized with respect to the scattered power of the uncloaked target, when a plane wave impinges from the vertical direction. The results of the cloak are plotted along with the calculated scattering for the reference behavior, i.e. an elliptical object shaped as $\partial \Omega^-$. Figure \ref{Figure7}(c) Oblique incidence with $45^o$ angel with respect to the horizontal direction. Figure \ref{Figure7}(d) Horizontal incidence.}
\label{Figure7}
\end{figure*}

\begin{figure*}[t]
 \centering
 \includegraphics[width=0.8\textwidth,angle=0] {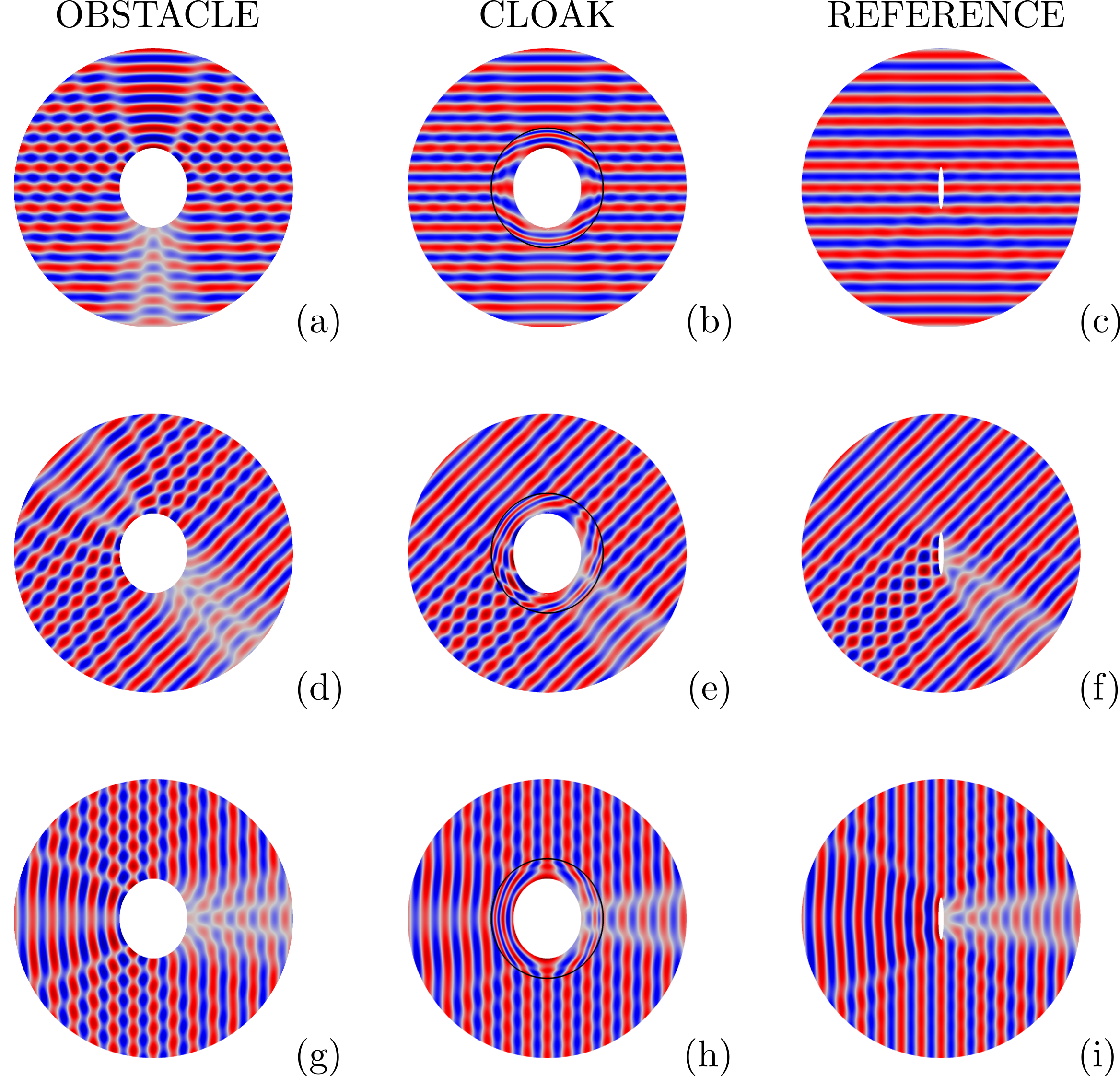}
 \caption{Total pressure field and pseudo-pressure calculated for the numerical assessment of performance illustrated in Figure \ref{Figure7}, for $V/\lambda=2$. from left to right the fields of the bare target without cloak, the cloaked obstacle and the reference obstacle are compared. From top to bottom the vetical incidence case, the oblique and the horziontal one are depicted, respectively.}
\label{Figure8}
\end{figure*}

To assess the difference in performance introduced by considering the transformation to be symmetric when instead the rotation is different from zero, numerical finite element simulations are conducted of the scattering of acoustic plane waves on a cloak designed for a target characterized by $H/V=0.85$ and $R_1=0.1R_2$. The outer ellipse is characterized by vertical semi-axis equal to 1.5 $V$. These parameters are set such that the maximum rotation reaches almost $50\;[^o]$, with an average $\mu_\alpha\approx 7\;[^o]$, as shown by Figure \ref{Figure7}(a). The goal is to assess if scattering reduction is still observed with such high rotation values. For a frequency range corresponding to $V/\lambda \in [0.5, 3]$ a plane wave is sent with different angles of incidence and the scattered power is integrated from the resulting scattering intensity to evaluate the \textit{total scattering cross section} 
\begin{equation}
    TSCS=\frac{W_{sc}}{I_{inc}}
\end{equation}
being $W_{sc}$ the scattered power and $I_{inc}$ the incident intensity. This procedure is repeated for the target provided with cloak, the bare target without cloak and a target shaped as the reference obstacle $\partial \Omega^-$ for comparison. The TSCS of the cloaked case and the reference obstacle are normalized with respect to the TSCS of the target obstacle without cloak, obtaining the reduction in scattered power, and plotted in Figure \ref{Figure7} for comparison. Figure \ref{Figure7}(b) refers to the case of vertical incidence, Figure \ref{Figure7}(c) to incidence at $\pi/4$ angle and Figure \ref{Figure7}(d) to horizontal incidence. As expected, the best case scenario is the vertical incidence, that still shows a broadband scattering reduction of approximately $80\%$ despite the neglected rotation, while the worst case scenario is the horizontal incidence, for which a $50\%$ reduction is calculated. The difference between the curve of the cloaked obstacle and the reference obstacle follows from the non-ideality of the solution, introduced by neglecting the rotation to obtain pure pentamode material properties. Figure \ref{Figure8} shows the calculated acoustic fields for $V/\lambda=2$.

\section{Microstructure alignment to principal directions of anisotropy}
Pentamode cloaks are usually implemented with solid microstructures after discretization of the continuous material distribution inside the cloak. Within each discretized domain, the specification of a constant density and a constant anisotropic elasticity tensor is obtained. Such material properties are then achieved by optimization of band diagrams in the long wavelength limit for unit cells showing the required symmetries, e.g. in 2D settings, centered rectangular lattices are usually employed because they posses the needed orthotropic elasticity tensor \cite{chen2015latticed} \cite{layman2013highly}. Each designed pentamode microstructure should then be "housed" within the respective part of the discretized cloak, aligning the principal directions of anisotropy with those prescribed by the transformation, while ensuring connectivity with the rest of the cloak.

This phase of the design can be greatly facilitated if the lines along which the main directions of anisotropy lie can be described by analytical functions: in this case, a direct mapping between the grid of the straight principal lines of the optimized rectangular lattice and the curvilinear grid of the principal directions in the cloak can indeed be set, which allows the accomodation of the microstructure in place to be mathematically performed at once. In the specific case at hand horizontal straight lines in a cartesian grid can be mapped to ellipses and vertical lines to hyperbolas by 
\begin{equation}
\begin{split}
    r=r(Y)\\
    \theta=\theta(X)\\
\end{split}
\end{equation}
where $r,\theta$ are elliptic coordinates describing points in the cloak, and $X,Y$ is the set of cartesian coordinates used in the design of the microstructure.\\

For illustrative purposes consider to know the optimized geometry of the microstructure of the centered rectangular lattice for a given portion of the cloak (Figure \ref{Figure5}): the four corners will be label with capital letters A-D, to easily identify their coordinates. A simple linear map as:
\begin{figure*}[t]
 \centering
 \includegraphics[width=\textwidth,angle=0] {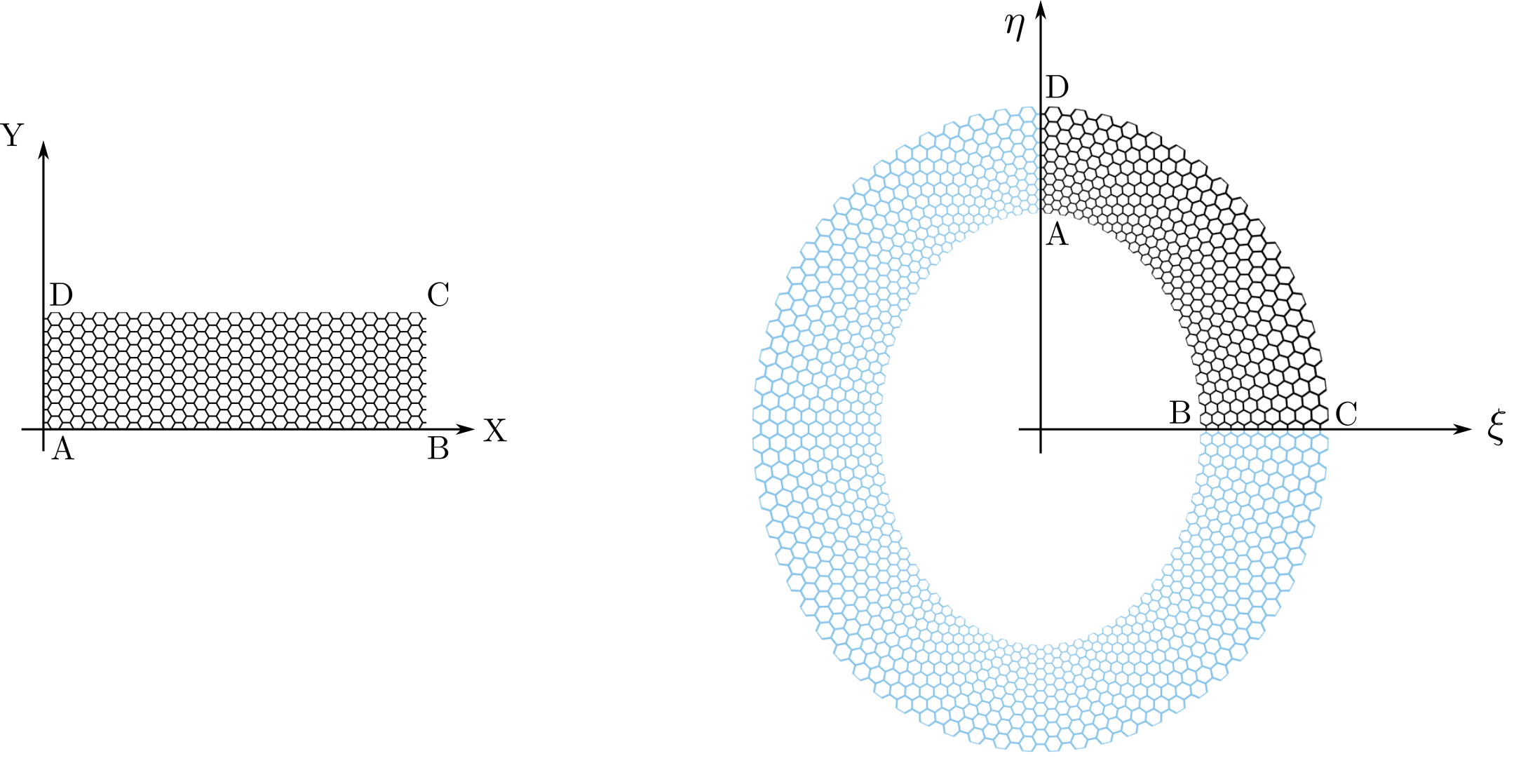}
 \caption{Deformation of a latticed microstructure to fit inside the geometry of the cloak, in such a way that principal directions of anisotropy remain aligned with coordinate lines of the elliptical system. Capital letters A, B, C, D are adopted for the four corners of the microstructure.}
\label{Figure9}
\end{figure*}
\begin{equation}
\begin{cases}
r=r_A+\displaystyle\frac{r_D-R_A}{Y_D-Y_A}(Y-Y_A)\\[10pt]
\theta=\theta_A+\displaystyle\frac{\theta_B-\theta_A}{X_B-X_A}(X-X_A)
\end{cases}
\label{microdef}
\end{equation}
will put the microstructure in place inside the portion of the cloak defined by the corresponding coordinates $r_i,\theta_i$, aligning principal directions of anisotropy to the coordinate lines of the elliptic system. The new cordinates of points in the cloak, according to a cartesian system can then be recovered as: 
\begin{equation}
\begin{cases}
\xi=a\sinh(r_A+\displaystyle\frac{r_D-r_A}{Y_D-Y_A}(Y-Y_A))\sin(\displaystyle\theta_A+\frac{\theta_B-\theta_A}{X_B-X_A}(X-X_A))\\[10pt]
\eta=a\cosh(r_A+\displaystyle\frac{r_D-r_A}{Y_D-Y_A}(Y-Y_A))\cos(\displaystyle\theta_A+\frac{\theta_B-\theta_A}{X_B-X_A}(X-X_A))
\end{cases}
\end{equation}
The underlying hypothesis is that after this procedure the geometry of the microstructure ideally remains unchanged, so that the previously optimized dispersion remains the same after the cloak is assembled. At this point we show that the important property of the elliptical system to have equal scaling factors can be exploited to easily guarantee that no distortion occurs during the transformation.
Note that the gradient of the deformation defined by Eq. \ref{microdef} is
\begin{equation}
    \F=\frac{r_D-R_A}{Y_D-Y_A}\g_r \otimes \E^Y+ \frac{\theta_B-\theta_A}{X_B-X_A}\g_\theta \otimes \E^X
\end{equation}
thus if $\displaystyle\frac{r_D-R_A}{Y_D-Y_A}=\displaystyle\frac{\theta_B-\theta_A}{X_B-X_A}$ or
\begin{equation}
    \frac{r_D-R_A}{\theta_B-\theta_A}=\frac{Y_D-Y_A}{X_B-X_A}
    \label{eqmarriage}
\end{equation}
since $|\g_r|=|\g_\theta|$ everywhere, it follows that
\begin{equation}
    \F=\V\R=\lambda_{iso}\R    
\end{equation}
with
\begin{equation}
    \begin{split}
        &\lambda_{iso}=\frac{r_D-R_A}{Y_D-Y_A}a\sqrt{\sinh^2(r)+\sin^2(\theta)}\\[10pt]
        &\R=\e_r\otimes\E^Y+\e_\theta \otimes \E^X
    \end{split}
\end{equation}
Thus, the transformation is represented everywhere by a rotation in combination with an isotropic expansion/contraction, i.e. it is a so-called conformal map that preserves angles and relative distances by points. Equation \ref{eqmarriage} can thus be used as a design rule to set up the discretization of the cloak together with the design of the associated unit cells so that they can retain their shape after adaptation to the target.

\section{Concluding Remarks}
In this manuscript the design problem of elliptical-shaped cloaks is tackled analytically, showing a strategy to write quasi-symmetric transformations adopting elliptic coordinates. The symmetry of the deformation gradient is the fundamental requirement for pure pentamode cloaking, which is crucial to obtain design that can be implemented with solid microstructures. A quantifiable approximation is introduced when considering the quasi-symmetric transformation for the design of pentamode cloaks with isotropic inertia, and can be related to the local rotation, which is analytically calculated.  The peculiarities of the approach illustrated in this manuscript are discussed and compared with respect to the usual axisymmetric cloaking, such as the sensitivity of performance with respect to the direction of incidence, for targets showing high ratios of vertical to horizontal semi-axis. Fully-coupled acusto-elastic numerical simulations of the scattering problem are performed to show the broadband nature of the performance of the solution, even in presence of non-negligible levels of approximation.

The property of the elliptic coordinate system having equal scale factors is exploited to underline the ease in setting transformations for specified distributions of stiffness properties in the cloak, given that the elasticity tensor depends on the mapping function alone. In addition, analytical ways of conformally accomodate previously optimized microstructures to the shape of the cloak to match the required principal directions of anisotropy are also discussed. In this way, we try to overcome some limits of the previously developed numerical approach for designing arbitrarily shaped pentamode cloaks, while at the same time this is paid with a restriction on the shape of the target that can be treated, putting itself in between the aforementioned numerical method and classical exact analytical solutions. Such restriction is justified by the relevance of elliptical shapes for underwater applications. In this respect, this work paves the way for experimental validation of the cloaking concept for non-axisymmetric targets in a feasible way.

\section{Acknowledgements}
The authors are grateful to the Italian Ministry of Defense for the financial support granted to this work through the PNRM "SUWIMM".

\appendix
\section{Elliptic coordinates}
\label{AppeA}
Recall the definition of elliptic coordinates:
\begin{equation}
\begin{cases}
x=a \sinh \mu \sin \nu\\
y=a \cosh \mu \cos \nu
\end{cases}
\end{equation}
From this relationship it is possible to obtain the basis vectors as:
\begin{equation}
\begin{cases}
\boldsymbol{g}_{\mu}=\displaystyle \frac{\partial x}{\partial \mu}\boldsymbol{e}_x + \frac{\partial y}{\partial \mu} \boldsymbol{e}_y= a \sin \nu \cosh \mu \boldsymbol{e}_x + a \cos \nu \sinh \mu \boldsymbol{e}_y \\
\boldsymbol{g}_{\nu}=\displaystyle \frac{\partial x}{\partial \nu}\boldsymbol{e}_x + \frac{\partial y}{\partial \nu} \boldsymbol{e}_y= a \sinh \mu \cos \nu \boldsymbol{e}_x - a \sin \nu \cosh \mu \boldsymbol{e}_y
\end{cases}
\end{equation}
Thus the metric tensor can be evaluated as:
\begin{equation}
[g_{ij}]=\begin{bmatrix}
a^2(\sinh^2 \mu + \sin^2 \nu)	& 	0\\
0 					&   a^2(\sinh^2 \mu + \sin^2 \nu)
\end{bmatrix}
\end{equation}
The contravariant metric tensor reads:
\begin{equation}
[g^{ij}]=\begin{bmatrix}
\displaystyle\frac{1}{a^2(\sinh^2 \mu + \sin^2 \nu)}		& 	0\\
0								&   \displaystyle\frac{1}{a^2(\sinh^2 \mu + \sin^2 \nu)}
\end{bmatrix}
\end{equation}
Contravariant base vectors are thus:
\begin{equation}
\begin{cases}
\boldsymbol{g}^{\mu}=\boldsymbol{g}_{\mu}\displaystyle\frac{1}{|\boldsymbol{g}_{\mu}|^2}\\
\boldsymbol{g}^{\nu}=\boldsymbol{g}_{\nu}\displaystyle\frac{1}{|\boldsymbol{g}_{\nu}|^2}
\end{cases}
\end{equation}

\bibliographystyle{elsarticle-num}
\bibliography{Mybibliography}

\end{document}